\documentclass{iau}
\usepackage{graphicx}

\title[Stellar energetic particles in young stars ] 
{Constraining the stellar energetic particle flux in young solar-like stars}

\author[Ch. Rab et al.]   
{Rab Ch.$^1$
 \and Padovani M.$^2$
 \and G\"udel M.$^3$
 \and Kamp I.$^1$
 \and Thi~W-F.$^4$
 \and Woitke P.$^5$}

\affiliation{$^1$Kapteyn Astronomical Institute, University of Groningen, P.O. Box 800, 9700 AV Groningen, The Netherlands, email: {\tt rab@astro.rug.nl}
\\[\affilskip]
$^2$University of Vienna, Dept. of Astrophysics, T\"urkenschanzstr. 17, 1180 Wien, Austria
\\[\affilskip]
$^3$INAF-Ossevatorio Astrofisico di Arcetri, Largo E. Fermi, 5 - 50125 Firenze, Italy
\\[\affilskip]
$^4$MPE, Giessenbachstrasse 1, 85748 Garching, Germany
\\[\affilskip]
SUPA, School of Physics \& Astronomy, University of St. Andrews, St. Andrews KY16 9SS, UK}
\pubyear{2019}
\volume{345}  
\jname{Origins: from the Protosun to the First Steps of Life}
\editors{Bruce G. Elmegreen, L. Viktor T\'oth, Manuel G\"udel, eds.}
\begin{document}
\maketitle
\begin{abstract}
Anomalies in the abundance measurements of short lived radionuclides in meteorites indicate that the protosolar nebulae was irradiated by a large number of energetic particles ($\mathrm{E\gtrsim10\,MeV}$), often called solar cosmic rays. The particle flux of the contemporary Sun cannot explain these anomalies, but, similar to \mbox{T Tauri} stars, the young Sun was more active and probably produced enough high energy particles. However, the stellar particle (SP) flux of young stars is essentially unknown. We model the impact of high-energy ionization sources on the chemistry of the circumstellar environment (disks and envelopes). The model includes X-ray radiative transfer and makes use of particle transport models to calculate the individual molecular hydrogen ionization rates. We study the impact on the chemistry via the ionization tracers HCO$^+$ and N$_2$H$^+$. We argue that spatially resolved observations of those molecules combined with detailed models allow for disentangling the contribution of the individual high-energy ionization sources and to put constraints on the SP flux in young stars.

\end{abstract}
\firstsection 
\section{Introduction \& Methods}
Our Sun can accelerate particles to energies similar to galactic cosmic rays ($\mathrm{E\gtrsim10\,MeV}$), hence this stellar energetic particles (SP) are also called solar cosmic rays. Abundance measurements of short-lived radionuclides in meteorites indicate that the protosolar nebulae experienced an about $10^5$ times higher SP flux than the Sun produces nowadays (see e.g. \cite{Feigelson2002,Gounelle2006}). Similar to T~Tauri stars (young solar analogues) the young Sun was likely more active and was probably  able to produce the SP flux required to explain the abundance anomalies measured in meteorites. This scenario is also supported by the observed high X-ray luminosities of T Tauri stars (\cite{Gudel2007}). However, contrary to X-rays,  the SP flux in T Tauri stars cannot be directly measured, and is not well constrained.

We present 2D radiation-thermo chemical models to study the impact of SPs on the ionization structure of the circumstellar environment. We use P{\small RO}D{\small I}M{\small O} ({PROtoplanetary DIsk MOdel, \cite{Woitke2009,Kamp2017}) to calculate the temperature structure, chemical abundances and to produce synthetic observables. We include the main $\mathrm{H_2}$ ionization sources: X-rays, galactic cosmic-rays (CR) and SPs. To calculate the $\mathrm{H_2}$ ionization rate we use X-ray radiative transfer (\cite{Rab2018}) and make use of particle transport models (\cite{Padovani2009}). This allows us to study the individual contributions of the ionization agents to the total ionization rate and on the chemistry of observable molecules such as HCO$^+$ and N$_2$H$^+$. 
\begin{figure}[t]
\begin{center}
\includegraphics[width=0.9\textwidth]{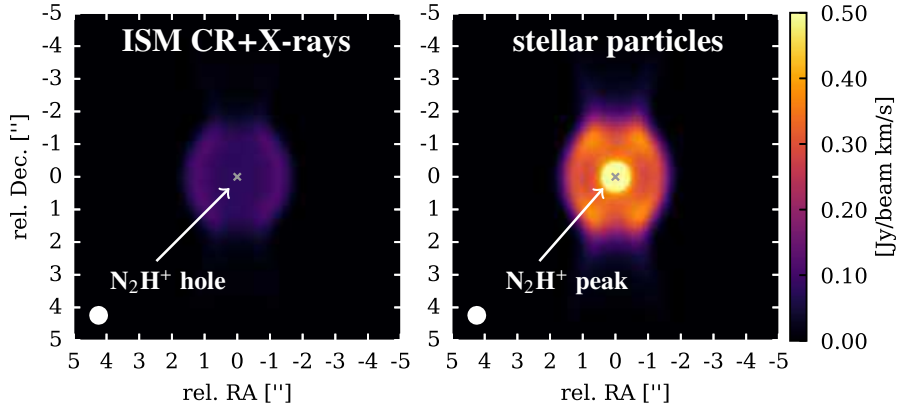}
\end{center}
\caption{Synthetic $\mathrm{N_2H^+} J=3-2$ integrated intensity maps for an envelope model (Class 0/I) without SP irradiation (left panel) and the same model with SPs (right panel).}
\label{fig1}
\end{figure}
\vspace*{-0.3cm}
\section{Results \& Conclusions}
In \cite{Rab2017}, we show that SPs can be the dominant $\mathrm{H_2}$ ionization source in the upper layers of a protoplanetary disk. Differently to X-rays and CRs, SPs cannot penetrate down to the midplane as they originate close to the star and are not scattered towards the midplane. With observations of two molecules, HCO$^+$, which traces the upper disk layers and N$_2$H$^+$, which traces deeper layers of the disk, it is therefore possible to measure the impact of SPs and to constrain the SP flux in young stars. 

Herschel observations of \cite{Ceccarelli2014} for the deeply embedded Class 0 protostar OMC-2~FIR~4 show very low HCO$^+$/N$_2$H$^+$ line ratios. They explained their observations with a strongly elevated $\mathrm{H_2}$ ionization rate of $10^{-14}-10^{-12}\,\mathrm{s^{-1}}$, produced by a high flux of SPs. We used the same approach as for our disk models but now for an envelope structure (see \cite{Rab2017a} for details) to simulate the conditions for embedded sources. We find that indeed a high SP flux is required to reproduce the low HCO$^+$/N$_2$H$^+$ ratios observed by \cite{Ceccarelli2014}. In Fig.~\ref{fig1}, we present ALMA simulations for this model showing that spatially resolved N$_2$H$^+$ observations should clearly show the signatures of SP ionization in embedded sources. We plan to also include different particle transport methods in our model (e.g. \cite{Rodgers2017}) to test if they produce yet other observable signatures.

\end{document}